\documentclass[a4paper]{jpconf}
\usepackage{graphicx}
\usepackage{iopams} 
\usepackage{graphics}
\usepackage{rotating}
\begin{document}
\title{Multistep shell model in the complex energy plane}

\author{Z.X. Xu, C. Qi, R. J. Liotta, R. Wyss}

\address{KTH, Alba Nova University Center, SE-10691 Stockholm, Sweden}

\ead{chongq@kth.se}

\begin{abstract}
We have adopted the multistep shell model in the complex energy
plane to study nuclear excitations occurring in the continuum part of
the spectrum. In this method  one proceeds by
solving the shell model equations in a successive manner. That is, in each
step one constructs the building blocks to be used in future steps.
We applied this formalism to analyze the unbound nuclei $^{12,13}$Li starting from the one-particle states in $^{10}$Li and two-particle states in $^{11}$Li. In the former case the excitations correspond to the motion of three
particles partitioned as the product of a one-particle and two-particle systems. The ground state of $^{12}$Li is thus calculated to be an antibound (virtual) state. In the four-particle system $^{13}$Li the states can be constructed as the coupling of two correlated pairs. We found that there is no bound or antibound state in $^{13}$Li.
\end{abstract}

\section{Introduction}

Experimental
facilities allow one nowadays to measure systems living a very short time. To describe
these processes one has
to consider the decaying character of the system. On the other hand, the theoretical study of unstable nuclei is a difficult undertaking. However, the system may be considered stationary
if it lives a long time. In this case the time dependence can be circumvented.
Of the various theories
that have been conceived to analyze unbound systems, we mention an extension
of the
shell model to the complex energy plane~\cite{idb02}. The basic assumption
of this theory is that resonances can be described in terms of states
lying in the complex energy plane. The real parts of the corresponding
energies are the positions of the resonances
while the imaginary parts are minus twice the corresponding widths, as it was
proposed by Gamow at the beginning of quantum mechanics~\cite{gam28}.
These complex states correspond to solutions of the Schr\"odinger equation with
outgoing boundary conditions. Details of the formalism and its application to two-nucleon systems can be found, e.g., in Refs.
\cite{mic02,cxsm,mic09}. 

Our aim is
to develop a suitable formalism to treat
unstable nuclei involving many valence nucleons in the continuum \cite{cxmsm}. This formalism is an extension of the shell model in the
complex energy plane~\cite{idb02}. The correlations induced by the pairing force acting upon
particles moving in decaying single-particle states is taken into
account by using the multistep shell model (MSM)~\cite{blo84}.
The formalism is presented in Section \ref{form}. Applications are in
Section \ref{appl} and a summary and conclusions are in Section~\ref{sumc}.

\section{The formalism}
\label{form}

\begin{figure}
  \includegraphics[width=\textwidth]{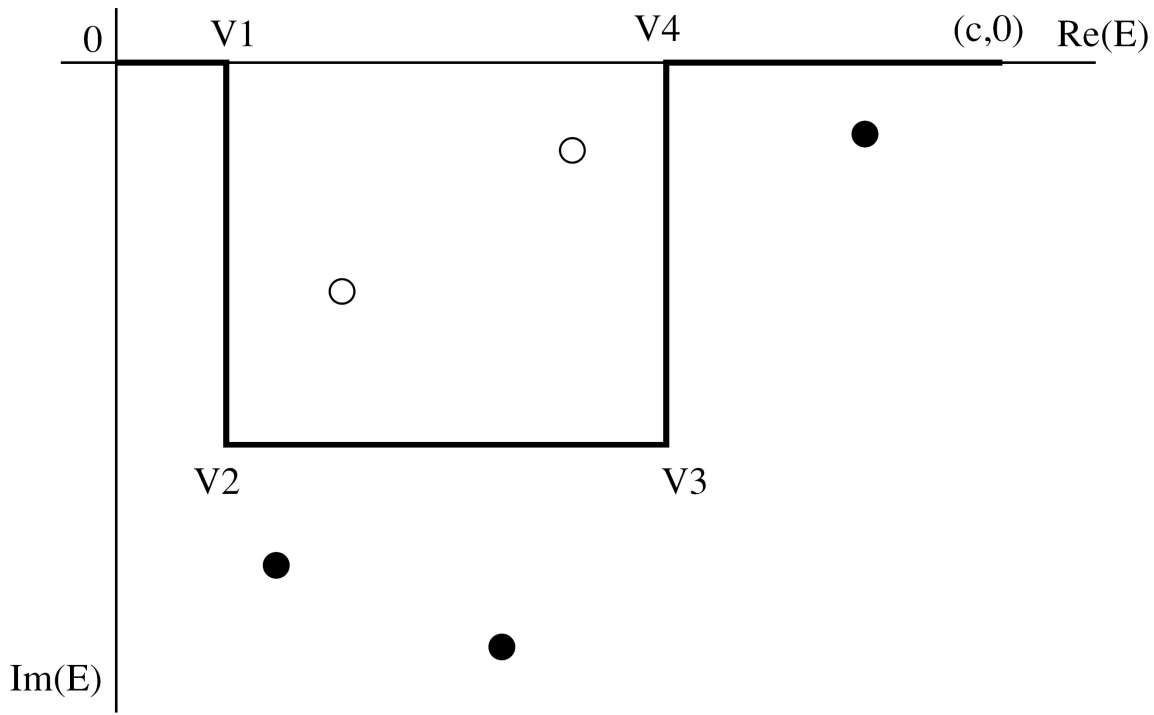}\\
  \caption{Integration contour $L^+$ in the complex energy plane. The open circles denote the resonances included in the sum of Eq. (\ref{comp}), while the solid circles are those excluded. The vertex $(c, 0)$ corresponds to the energy cutoff point $c$.}\label{Cauchy}
\end{figure}

The eigenstates of a central potential obtained as outgoing
solutions of the Schr\"odinger equation can be used to
express the Dirac $\delta$-function as \cite{b68}
\begin{equation}\label{eq:delb}
\delta(r-r')=\sum_n w_n(r) w_n(r') + \int_{L^+} dE u(r,E) u(r',E),
\end{equation}
where the sum runs over all the bound and antibound states plus the complex
states (resonances) which lie between the real energy axis and the integration contour
$L^+$ (see, Fig. \ref{Cauchy}). The antibound states are virtual states with negative scattering length. The wave function of a state $n$ in these discrete set is
$w_n(r)$ and  $u(r,E)$ is the scattering function at energy $E$.

Discretizing the integral of Eq. (\ref{eq:delb}) one obtains
the set of orthonormal vectors $\vert \varphi_j\rangle$
forming the Berggren representation \cite{lio96}. Since this discretization
provides an approximate value of the integral, the Berggren vectors fulfill
the relation $I\approx\sum_j \vert \varphi_j\rangle \langle \varphi_j\vert$. One has,
\begin{equation}\label{comp}
    \int_{L^+}dE u(r,E)u(r',E)=\sum_p h_p u(r,E_p)u(r',E_p),
\end{equation}
where the quantities $E_p$ and $h_p$ are defined by the procedure that uses to perform the integration. Therefore the orthonormal vectors $|\varphi_j\rangle$ are given by the set of bound, antibound and Gamow states, i.e., $\langle r|\varphi_n\rangle=\{w_n(r,E_n)\}$, and the discretized scattering states, i.e., $\langle r|\varphi_p\rangle=\{\sqrt{h_p}u_p(r,E_p)\}$. In Ref. \cite{lio96} it was also found that few discretized scattering states in the basis are enough to obtain convergence.
Using the Berggren representation one readily gets the
two-particle shell-model equations in the complex energy plane (CXSM)
\cite{cxsm}, i.e.,
\begin{equation}\label{eq:sme}
(W(\alpha_2)-\epsilon_i-\epsilon_j)X(ij;\alpha_2)=
\sum_{k\leq l}\langle\tilde k\tilde l;\alpha_2\vert V\vert ij;\alpha_2\rangle X(kl;\alpha_2),
\end{equation}
where $V$ is the residual interaction and $i,j,k,l$ label single-particle states. The two-particle states are labeled by
$\alpha_2$. $\epsilon$ and $W$ denote the energy of the single-particle and two-particle states, respectively. The tilde in the interaction matrix
element denotes mirror states so that in
the corresponding radial integral there is not any complex conjugate,
as required by the Berggren metric.

The two-particle
wave function is given by
\begin{equation}\label{eq:wfsq}
\vert \alpha_2\rangle=P^+(\alpha_2)\vert 0\rangle,
\end{equation}
where the two-particle creation operator is
\begin{equation}\label{eq:wfsq2}
P^+(\alpha_2)=\sum_{i\leq j}X(ij;\alpha_2)
\frac{(c^+_ic^+_j)_{\lambda_{\alpha_2}}}{\sqrt{1+\delta_{ij}}},
\end{equation}
and $\lambda_{\alpha_2}$ is the angular momentum
of the two-particle state.

\subsection{Separable interaction}\label{separable}

For a separable interaction the matrix can be written as
\begin{equation}
    \langle\widetilde{kl};\alpha|V|ij;\alpha\rangle=-G_\alpha f_\alpha(kl)f_\alpha(ij),
\end{equation}
where $G_\alpha$ is the interaction strength and $f_\alpha(ij)$ is the matrix element of the field defining the interaction. The shell model equation becomes
\begin{equation}\label{sep1}
    -\frac{X(ij;\alpha)}{G_\alpha}=\sum_{k\leqslant l}\frac{f_\alpha(kl)f_\alpha(ij)X(kl;\alpha)}{(\omega_\alpha-\epsilon_i-\epsilon_j)}.
\end{equation}
Multiplying $\sum_{i\leqslant j}f_\alpha(ij)$ on both sides of Eq. (\ref{sep1}) one gets the so-called dispersion relation
\begin{equation}
    -\frac{1}{G_\alpha}=\sum_{i\leqslant j}\frac{f^2_\alpha(ij)}{\omega_\alpha-\epsilon_i-\epsilon_j}.
\end{equation}
The two-particle wave function amplitudes are given by
\begin{equation}
    X(ij;\alpha)=N_\alpha\frac{f_\alpha(ij)}{\omega_\alpha-\epsilon_i-\epsilon_j},
\end{equation}
where $N_\alpha$ is the normalization constant determined by requiring
\begin{equation}
    \sum_{i\leqslant j}X^2(ij;\alpha)=1.
\end{equation}

\subsection{The Multistep Shell Model Method}
\label{sec:msm}
The Multistep Shell Model Method (MSM) solves the
shell model equations in several steps. In the first step the
single-particle representation is chosen. In the second step the energies
and wave functions of the two-particle system are evaluated with a given
two-particle interaction. The three-particle states are evaluated in terms
of a basis consisting of the tensorial product of the one- and two-particle
states previously obtained.  The MSM basis is overcomplete and non-orthogonal.
To correct this one needs to evaluate the overlap matrix among the basis
states also. A general description of the formalism is in Ref. \cite{lio82}.
The particular system that is of our interest here, i.e., the
three-particle case, can be found in Ref. \cite{blo84}.
Below we
refer to this formalism as CXMSM \cite{cxmsm}.

The three-particle energies $W(\alpha_{3})$ are given by \cite{blo84}
\begin{eqnarray}\label{TDA3}
\nonumber    (W(\alpha_{3})-\varepsilon_{i}-W(\alpha_{2}))\langle\alpha_{3}|(c^{+}_{i}P^{+}(\alpha_{2}))_{\alpha_{3}}|0\rangle\\
    =\sum_{j\beta_{2}}\left\{\sum_{k}(W(\beta_{2})-\varepsilon_{i}-\varepsilon_{k})A(i\alpha_{2},j\beta_{2};k)\right\}
    \langle\alpha_{3}|(c^{+}_{j}P^{+}(\beta_{2}))_{\alpha_{3}}|0\rangle,
\end{eqnarray}
where
\begin{equation}
    A(i\alpha_{2},j\beta_{2};k)=\hat{\alpha}_{2}\hat{\beta}_{2}\left\{\begin{array}{ccc}
                                                                        i & k & \beta_{2} \\
                                                                        j & \alpha_{3} & \alpha_{2}
                                                                      \end{array}
    \right\}Y(kj;\alpha_{2})Y(ki;\beta_{2}),
\end{equation}
and
\begin{equation}
    Y(ij;\alpha_{2})=(1+\delta(i,j))^{1/2}X(ij;\alpha_{2}).
\end{equation}
The rest of the notation is standard.

The matrix defined in Eq. (\ref{TDA3}) is not hermitian and the dimension 
may be larger than the corresponding shell-model dimension. This is due to the violations of the Pauli principle as well as over-counting of
states in the CXMSM basis. Therefore the direct diagonalization of Eq. (\ref{TDA3}) is not convenient.
One needs to calculate the overlap matrix in order to transform the CXMSM basis into an orthonormal set.
In this three-particle case the overlap matrix is
\begin{equation}\label{Overlap}
    \left\langle0|(c^{+}_{i}P^{+}(\alpha_{2}))^{\dag}_{\alpha_{3}}(c^{+}_{j}P^{+}(\beta_{2}))_{\alpha_{3}}|0\right\rangle
    =\delta_{ij}\delta_{\alpha_{2}\beta_{2}}+\sum_{k}A(i\alpha_{2},j\beta_{2};k).
\end{equation}

Using this matrix (\ref{Overlap}) one can transform the matrix determined by Eq. (\ref{TDA3})
into a hermitian matrix $T$ which has the right dimension. The diagonalization of $T$ provides the three-particle energies. The
corresponding wave function amplitudes can be readily evaluated to obtain
\begin{eqnarray}
    |\alpha_{3}\rangle&=&P^{+}(\alpha_{3})|0\rangle,\\
    P^{+}(\alpha_{3})&=&\sum_{i\alpha_{2}}X(i\alpha_{2};\alpha_{3})(c^{+}_{i}P^{+}(\alpha_{2}))_{\alpha_{3}},
\end{eqnarray}
where $P^{+}(\alpha_{3})$ is the three-particle creation operator.
It has to be pointed out that in cases where the basis is overcomplete the amplitudes
$X$ are not well defined. But this is no hinder to evaluate the physical quantities.
For details see Ref. \cite{blo84}.

The CXMSM allows one
to choose in the basis states a limited number of excitations. This is because in the continuum
the vast majority of basis states consists of scattering functions. That is, the
majority of the two-particle states provided by the CXSM are complex
states which form a part of the continuum background. Only a few of those
calculated states correspond to physically meaningful
resonances, i.e., resonances which can be observed. Below we call a ``resonance" only to a complex state which
is meaningful. These resonances  are mainly built
upon single-particle states which are either bound or narrow resonances.
Yet, one cannot ignore the continuum when evaluating the
resonances. The continuum configurations in the resonance wave function are
small but many, and they affect the two-particle resonance significantly
\cite{cxsm}. 
Therefore, the great advantage of the CXMSM is that one can include in the basis only
two-particle resonances, while neglecting the background continuum states,
which form the vast majority of complex two-particle states.

\subsection{The neutron-proton correlation}
The influence of the neutron-proton correlation can be considered in a straightforward way in our CXMSM. For example, we consider a system with three neutrons and one proton , say $^{12}$Li by assuming $^8$He as the inert core \cite{cxmsm}. With the basis denoted as
\begin{equation}
|p\alpha_3;\alpha_4\rangle=(c^+_p P^+(\alpha_3))_{\alpha_4}|0\rangle,
\end{equation}
where $p$ labels the proton state and
$\alpha_4$ are the three-neutron
one-proton state,  the four-particle energies $W(\alpha_4$) in $^{12}$Li are
given by
\begin{eqnarray}\label{TDA4}
\nonumber (W(\alpha_4)-\varepsilon_p-W(\alpha_3))\langle\alpha_4|(c^+_pP^+(\alpha_3))_{\alpha_4}|0\rangle\\
    =\sum_{q\beta_3}\left\{\sum_{kl\lambda\alpha_2}\langle pk;\lambda|V|ql;\lambda\rangle B_1+\sum_{ijkl\lambda\alpha_2\beta_2}\langle pi;\lambda|V|ql;\lambda\rangle B_2\right\}\langle\alpha_4|(c^+_qP^+(\beta_3))_{\alpha_4}|0\rangle,
\end{eqnarray}
where,
\begin{eqnarray}
    B_1=(-1)^{p+q+k+l}X(k\alpha_2;\alpha_3)F(l\alpha_2;\beta_3)\hat{\alpha}_3\hat{\beta}_3\hat{\lambda}^2\left\{
                                                                  \begin{array}{ccc}
                                                                    p & k & \lambda \\
                                                                    \alpha_2 & \alpha_4 & \alpha_3 \\
                                                                  \end{array}
                                                                \right\}\left\{
                                                                  \begin{array}{ccc}
                                                                    q & l & \lambda \\
                                                                    \alpha_2 & \alpha_4 & \beta_3 \\
                                                                  \end{array}
                                                                \right\},
\end{eqnarray}
and
\begin{eqnarray}
\nonumber    B_2=(-1)^{p+q+i+l}Y(ji;\alpha_2)Y(jk;\beta_2)X(k\alpha_2;\alpha_3)F(l\beta_2;\beta_3)\\
        \times\hat{\alpha}_2\hat{\alpha}_3\hat{\beta}_2\hat{\beta}_3\hat{\lambda}^2\left\{
                                                                                             \begin{array}{ccc}
                                                                                               p & i & \lambda \\
                                                                                               \beta_2 & \alpha_4 & \alpha_3 \\
                                                                                             \end{array}
                                                                                           \right\}\left\{
                                                                                             \begin{array}{ccc}
                                                                                               q & l & \lambda \\
                                                                                               \beta_2 & \alpha_4 & \beta_3 \\
                                                                                             \end{array}
                                                                                           \right\}\left\{
                                                                                             \begin{array}{ccc}
                                                                                               i & j & \alpha_2 \\
                                                                                               k & \alpha_3 & \beta_2 \\
                                                                                             \end{array}
                                                                                           \right\}.
\end{eqnarray}
Here $p$ and $q$ label proton states, while $i,j,k,l$ label neutron states.
$\langle pk;\lambda|V|ql;\lambda\rangle$ denotes the corresponding proton-neutron interaction matrix element. The wave function amplitudes $X$ and the projected quantities
$F$ have been evaluated in previous steps
of the CXMSM.
Notice that in this case the overlap matrix is the unit matrix, i.e.,
\begin{equation}
\langle 0|(c^+_{p'} P^+(\alpha'_3))^\dag_{\alpha_4}(c^+_p P^+(\alpha_3))_{\alpha_4}
|0\rangle =\delta_{pp'}\delta_{\alpha_3\alpha'_3}.
\end{equation}

\subsection{Four-particle states}
Systems with four like particles can be described in a correlated two-particle basis as
\begin{equation}
    |\alpha_2\beta_2;\alpha_4\rangle=\big(P^\dag(\alpha_2)P^\dag(\beta_2)\big)_{\alpha_4}|0\rangle,
\end{equation}
where $\alpha_2\leqslant\beta_2$. One can obtain the four-particle MSM equation as
\begin{eqnarray}\label{4pt}
 \nonumber   \big(W(\alpha_4)-W(\alpha_2)-W(\beta_2)\big)\langle\alpha_4|\big(P^\dag(\alpha_2)P^\dag(\beta_2)\big)_{\alpha_4}|0\rangle
        =-\sum_{\gamma_2\leqslant\delta_2}\sum_{ijkl}(1+\delta_{\gamma_2\delta_2})^{-1}\\
\times\big(W(\gamma_2)+W(\delta_2)-\epsilon_i-\epsilon_j-\epsilon_k-\epsilon_l\big) B(ijkl,\alpha_2\beta_2\gamma_2\delta_2;\alpha_4)\langle\alpha_4|\big(P^\dag(\gamma_2)P^\dag(\delta_2)\big)_{\alpha_4}|0\rangle,
\end{eqnarray}
where
\begin{eqnarray}
    B(ijkl,\alpha_2\beta_2\gamma_2\delta_2;\alpha_4)=\hat{\alpha}_2\hat{\beta}_2\hat{\gamma}_2\hat{\delta}_2 Y(ij;\alpha_2)Y(kl;\beta_2)Y(ik;\gamma_2)Y(jl;\delta_2)\left\{
                                                                       \begin{array}{ccc}
                                                                         i & j & \alpha_2 \\
                                                                         k & l & \beta_2 \\
                                                                         \gamma_2 & \delta_2 & \alpha_4 \\
                                                                       \end{array}
                                                                     \right\}.
\end{eqnarray}

The four-particle overlap matrix can be obtained in a similar way as
\begin{eqnarray}\label{4pto}
\nonumber    \langle0|\big(P^\dag(\alpha_2)P^\dag(\beta_2)\big)^\dag_{\alpha_4}\big(P^\dag(\gamma_2)P^\dag(\delta_2)\big)_{\alpha_4}|0\rangle=\delta_{\alpha_2\gamma_2}\delta_{\beta_2\delta_2}\\
        +(-1)^{\alpha_2+\beta_2-\alpha_4}\delta_{\alpha_2\delta_2}\delta_{\beta_2\gamma_2}-\sum_{ijkl}B(ijkl;\alpha_2\beta_2\gamma_2\delta_2).
\end{eqnarray}

To restore the Pauli principle, one can use the Schmidt procedure to generate a set of orthonormal basis $|u_i\rangle$ (which has the right dimensions) from the overlap matrix. The overlap matrix is thus transformed into an identity matrix
\begin{equation}
    \langle u_m|u_n\rangle=I_{mn}=\sum_{ij}\xi_m(i)O(i,j)\xi_n(j),
\end{equation}
where $O(i,j)$ is the overlap matrix, and the wave function in the Schmidt basis is
\begin{equation}\label{wfu}
    |u_m\rangle=\sum_i \xi_m(i) |i\rangle.
\end{equation}
One can transform the dynamical matrix into a hermitian matrix $T$ in the right dimensions if all possible basis states are included. The dynamical matrix $M(ij)$ in the MSM basis is given by
\begin{equation}
    W(\alpha)\langle i|\alpha\rangle=\sum_j M(i,j)\langle j|\alpha\rangle.
\end{equation}
To transform it into the orthonormal basis, one can use Eq. (\ref{wfu}) and the identity equation. One thus obtains
\begin{eqnarray}\label{T-m}
    W(\alpha)\sum_i \xi_m(i)\langle i|\alpha\rangle&=&\sum_i\xi_m(i)\sum_j M(i,j)\sum_n\langle j|u_n\rangle\langle u_n|\alpha\rangle,\\
    W(\alpha)\langle u_m|\alpha\rangle&=&\sum_n\bigg\{\sum_{ij}\xi_m(i)M(i,j)\sum_k\xi_n(k)\langle j|k\rangle\bigg\}\langle u_n|\alpha\rangle.
\end{eqnarray}
The expression in the bracket is therefore the hermitian matrix $T$, which can be written as
\begin{equation}
    T(m,n)=\sum_{ijk}\xi_m(i)M(i,j)O(j,k)\xi_n(k).
\end{equation}

\section{Applications}
\label{appl}
Using the Berggren single-particle representation described above, we will
evaluate the complex energies and wave functions of the unbound nuclei $^{12,13}$Li using the MSM basis states
consisting of the Berggren one-particular states, which are states in
$^{10}$Li, times the two-particle states corresponding to $^{11}$Li. 
The spectrum of $^{11}$Li was already evaluated within the CXSM including
antibound states~\cite{ant}. Here we will repeat that calculation in order to
determine the two-particle states to be used in the calculation of the
three- and four-particle systems.

To define the Berggren single-particle representation we still have to
choose the integration contour ${L^+}$ (see Eq. (\ref{eq:delb})).
The valence shells are the low lying resonances $0p_{1/2}$ at
(0.195,-0.047) MeV and $0d_{5/2}$ at (2.731, -0.545) MeV.
Besides, the state $1s_{1/2}$ appears as an
antibound state. To include in the representation the antibound $1s_{1/2}$ state as well as the
Gamow resonances $0p_{1/2}$ and $0d_{5/2}$ we will use two different contours.
The number of points on each contour defines the energies of the scattering
functions in the Berggren representation, i.e., the number of basis states
corresponding to the continuum background. This number is not uniformity
distributed, since in segments of the contour which are close to
the antibound state or to a resonance the scattering functions increase
strongly.

\begin{figure}[htdp]
\begin{center}
\includegraphics[scale=.750]{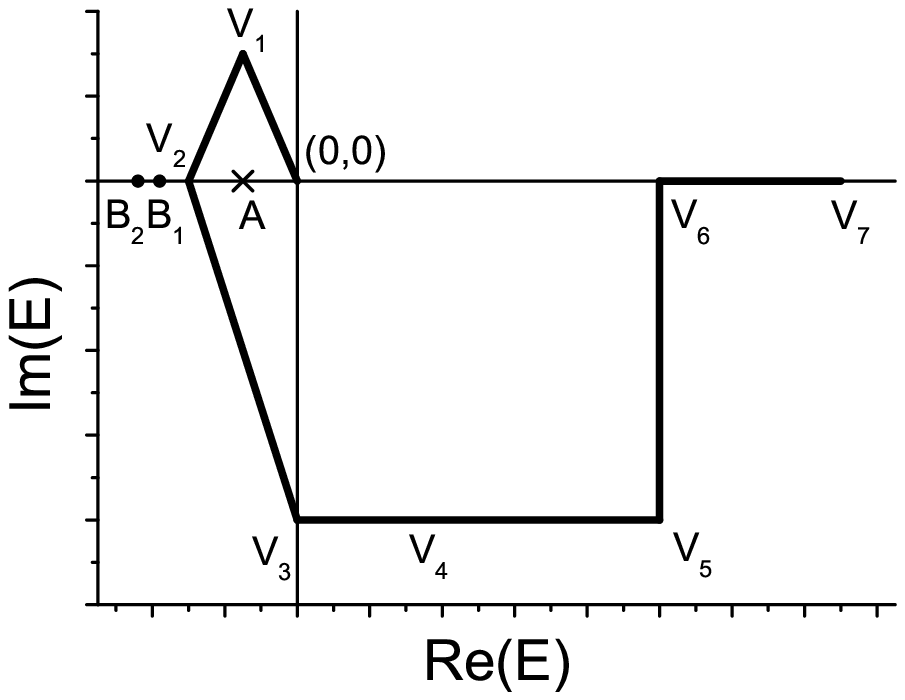} \includegraphics[scale=.750]{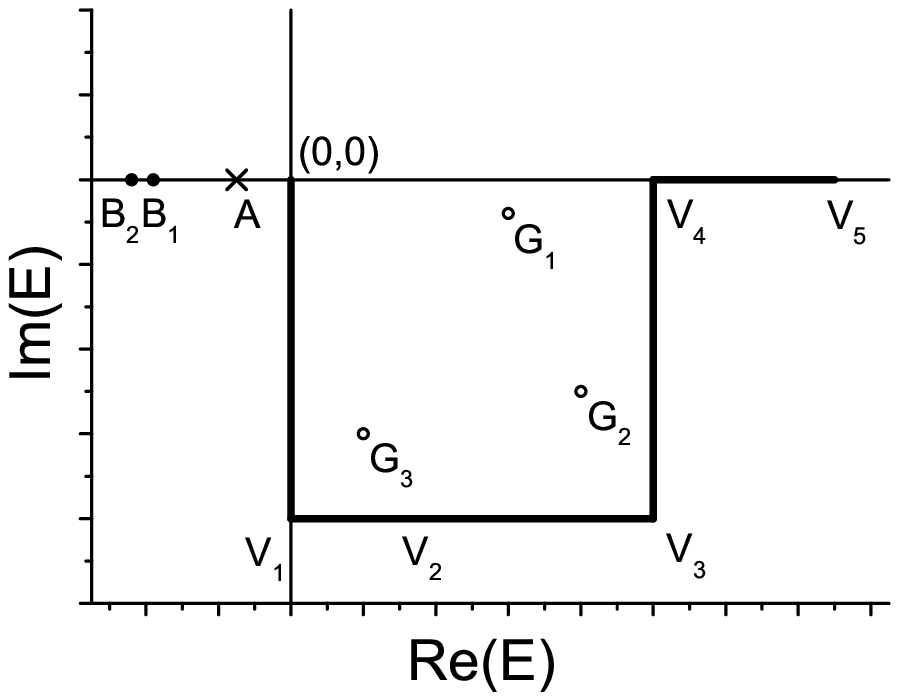}\\
\caption{Left: Contour used to include the antibound state (see, also, Ref.~\cite{idb04}). The points
$B_i$ denote bound state energies while
$A$ denotes the antiboud state. Right: Contour used to include the Gamow resonances represented by the
points $G_i$.}
\label{cont}
\end{center}
\end{figure}

The dynamics of  $^{11}$Li is determined by the pairing
force acting upon the two neutrons coupled to a state $0^+$, which behaves
as a normal even-even ground state \cite{esb97,ant}.
Besides the energy, this state has been measured to have an angular momentum
contain  of about 60 \% of  $s$-waves and 40 \% of $p$-waves,
although small components of
other angular momenta are not excluded.

We will use a separable interaction as in Ref. \cite{ant}. The
strength $G_{\lambda_2}$, corresponding to the states with
angular momentum $\lambda_2$ and parity $(-1)^{\lambda_2}$, will be
determined by fitting the experimental energy of the lowest of these states,
as usual. It is worthwhile to point out that $G_{\lambda_2}$
defines the Hamiltonian and, therefore, is a real quantity. 
The energies are thus obtained by solving the corresponding
dispersion relation.
The two-particle wave function amplitudes are given by
\begin{equation}
\label{eq:tpwf}
 X(ij;\alpha_2) = N_{\alpha_2} \frac{f(ij,\alpha_2)}{\omega_{\alpha_2}-
(\epsilon_i + \epsilon_j)},
\end{equation}
where $f(ij,\alpha_2)$ is the single particle matrix element of the field
defining the separable interaction and
$N_{\alpha_2}$ is the normalization constant determined by the condition
$\sum_{i \le j} X(ij;{\alpha_2})^2 = 1$.

Due to the large number of scattering states included in the single-particle representation the dimension of
three-particle basis is also large. The scattering states are needed in
order to describe these unstable states.
In the calculations we took into account all the possibilities described
above regarding the energies of the single-particle state $0p_{1/2}$ as well
as the binding energy of the state $^{11}$Li(gs). 

\begin{figure}[htdp]
\begin{center}
\vspace{0.9cm}
\includegraphics[scale=0.50]{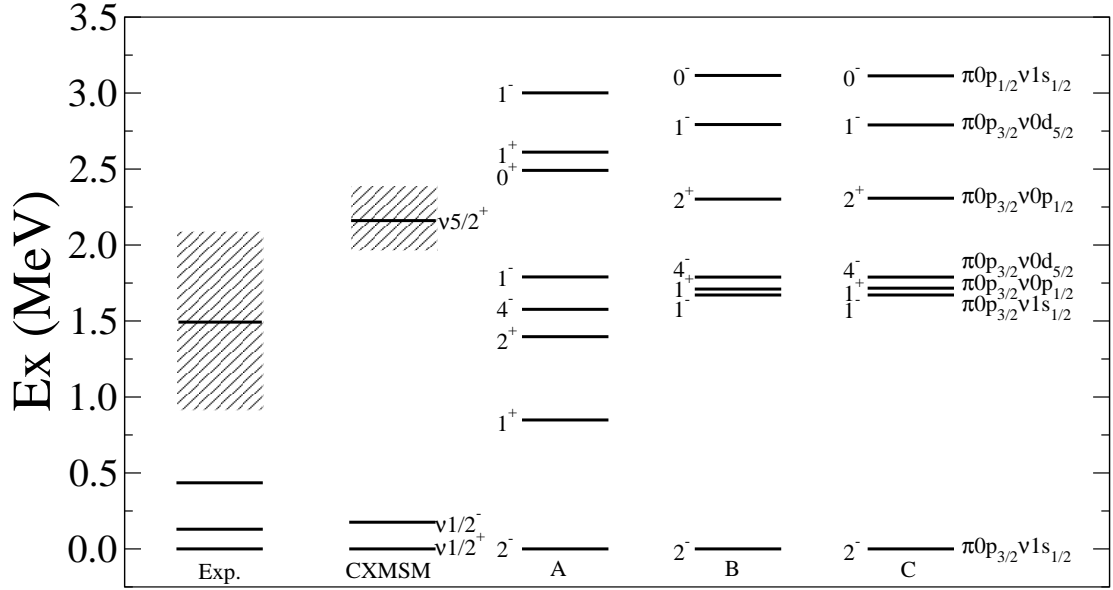}\\
\end{center}
\caption{ Experimental level scheme in $^{12}$Li. The three lowest levels are 
from \cite{hal10}, while the one at 1.5 MeV is from \cite{pat10}. In the second
column are the three-neutron CXMSM results.
In the columns A-C are the shell-model calculations corresponding to different truncation
schemes: A) Maximum of 1p excitation from $p$ to $sd$ shell, B) maximum of 3p excitations and
C) full $psd$ space. Dashed lines indicate the widths of the resonances.}
\label{smen}
\end{figure}

With the single-particle states and the two-particle states $^{11}$Li(gs) and 
$^{11}$Li($2^+_1$) discussed above, we formed all the possible
three-particle basis states. We found that the only physically relevant
states are those which are mainly determined by the bound state $^{11}$Li($0^+_1$).
The corresponding spins and parities are 
$1/2^+$, $1/2^-$ and $5/2^+$. States like $3/2^+$, which arises from the
CXMSM configuration $|1s_{1/2}\otimes 2^+_1;3/2^+\rangle$, is not a meaningful
state. The corresponding calculated energies are shown in Fig. \ref{smen}.

In Ref. \cite{hal10} it was also found that $^{12}$Li(gs) is an antibound state but, in addition, two other low-lying
states were observed at 0.250 MeV and 0.555 MeV by using
two-proton removal reactions. In this case the $0p_{3/2}$ proton in
the core may interfere with the neutron excitations evaluated above.
In particular, the antibound $1/2^+$ ground state would provide, through the proton
excitation, a state $1^-$ and a $2^-$. This is the situation
encountered in the shell model calculation \cite{qi08}. 
One sees in this Figure that the full calculation predicts all excited states
to lie well above the corresponding experimental values. It is worthwhile to
point out that the calculated states exhibit rather pure shell model
configurations. For instance the states $2^-_1$ (ground state) and $1^-_1$
are mainly composed of the configuration
$|\pi \left[ 0p_{3/2}\right]\nu \left[ (0p_{1/2})^2 1s_{1/2}\right]\rangle$.
This does not fully agree with our CXMSM calculation, since in our case this
wave function is mainly of the form 
$|1s_{1/2}\otimes ^{11}{\rm Li(gs)}\rangle$. This differs
from the shell model case in two ways. First, the state $^{11}$Li(gs)
contains nearly as much of $1s_{1/2}$ as of $0p_{1/2}$. Second the continuum
states contribute much in the building up of the antibound $^{12}$Li(gs)
wave function, as discussed above. In our representation it is
straightforward to discern the antibound character of this state, which is
not the case when using harmonic oscillator bases.

\begin{table}
  \centering
  \caption{Calculated energies (in MeV) of the four-particle states in $^{13}$Li corresponding to different single-particle $\epsilon_{p_{1/2}}$ energies. The two-particle states ${W(0^+_1)}=-0.295$MeV, ${W(1^-_1)}=(0.084,-0.002)$ and ${W(2^+_1)}=(2.300,-0.372)$MeV are included in the calculation.}\label{Li13}
  \vspace{0.2cm}
\begin{tabular}{cccc}
  \hline\hline
        &$\epsilon_{p_{1/2}}=(0.195,-0.047)$ &  $\epsilon_{p_{1/2}}=(0.563,-0.252)$ \\\hline
    $0^+$& (0.868,-0.059) & (1.505,-0.041)  \\
         & (5.127,-0.964) & (5.274,-1.009)  \\
    $1^-$& (0.836,-0.117) & (1.722,-0.166)  \\
         & (3.244,-0.593) & (3.921,-0.721)  \\
    $2^+$& (0.715,-0.114) & (1.802,-0.257)  \\
         & (2.907,-0.445) & (3.373,-0.478)  \\
         & (5.131,-0.910) & (5.205,-0.966)  \\
    $3^-$& (2.541,-0.391) & (2.674,-0.413)  \\
    $4^+$& (5.715,-1.119) & (5.715,-1.119)  \\
  \hline\hline
  \end{tabular}
\end{table}

\begin{table}
  \centering
  \caption{Same as Table \ref{Li13} but for ${W(0^+)}=-0.369$MeV.} \label{Li13s}
  \vspace{0.2cm}
\begin{tabular}{cccc}
  \hline\hline
          &$\epsilon_{p_{1/2}}=(0.195,-0.047)$ &  $\epsilon_{p_{1/2}}=(0.563,-0.252)$ \\\hline
    $0^+$& (0.855,-0.057) & (1.527,-0.033)  \\
         & (5.142,-0.970) & (5.303,-1.020)  \\
    $1^-$& (0.828,-0.122) & (1.744,-0.169)  \\ 
         & (3.244,-0.593) & (3.921,-0.721)  \\ 
    $2^+$& (0.715,-0.114) & (1.802,-0.257)  \\ 
         & (2.903,-0.450) & (3.367,-0.477)  \\ 
         & (5.137,-0.916) & (5.213,-0.973)  \\ 
    $3^-$& (2.541,-0.391) & (2.674,-0.413)  \\ 
    $4^+$& (5.715,-1.119) & (5.715,-1.119)  \\ 
  \hline\hline
  \end{tabular}
\end{table}

Using the two-particle states of $^{11}$Li within the Berggren single-particle representation, we can calculate the four-particle system $^{13}$Li. The four-particle CXMSM basis is constructed by two correlated two-particle states. The two-particle states we included here are the $^{11}$Li($0^+_1$), $^{11}$Li($1^-_1$) and $^{11}$Li($2^+_1$), which are the same as the three-particle case. In this basis set we formed the symmetric Hamiltonian matrix by evaluating the dynamical matrix Eq. (\ref{4pt}) and the overlap Eq. (\ref{4pto}). 

The only four-particle states we can get are $0^+$, $1^-$, $2^+$, $3^-$ and $4^+$ states. The calculated results refer to different single-particle $\epsilon_{p_{1/2}}$ and two-particle $^{11}$Li(gs) energies are all listed in Tables \ref{Li13} and \ref{Li13s}. We found that there is no bound or antibound state in $^{13}$Li, which agrees with the experiment that the $^{13}$Li is unbound \cite{aks08}. 
The ground state is a resonance at around $1$ MeV, however the configuration and the energy strongly depend on which single-particle energy $\epsilon_{p_{1/2}}$ we choose. 
For $\epsilon_{p_{1/2}}=0.195$MeV case, the ground state is $|^{11}$Li($1^-_1)\otimes^{11}$Li($1^-_1$);$2^+\rangle$ at the energy of $(0.715,-0.114)$MeV, 
while for $\epsilon_{p_{1/2}}=0.563$MeV case, the ground state is $0^+$ which at the energy of $(1.505,-0.041)$MeV and 
with the components $|^{11}$Li($1^-_1)\otimes^{11}$Li($1^-_1);0^+\rangle$ and 
$|^{11}$Li($0^+_1)\otimes^{11}$Li($0^+_1);0^+\rangle$ about half and half. Comparing between tables one can find that the four-particle states are not much affected by the energy of $^{11}$Li(gs).

In our calculation we only included the resonances of $^{11}$Li which is considered physically meaningful for building the whole four-particle basis. However as we mentioned before, there are lots of other continuum states which belong to the background and have a large component of the poles. Although these states are not in themselves physically meaningful resonances, they might have some influence on the spectrum of $^{13}$Li. Therefore for a more precise calculation, one should include some of these two-particle states. 

\section{Summary and conclusions}
\label{sumc}

We have studied excitations occurring in the continuum part of
the nuclear spectrum which are at the limit of what can be observed within
present experimental facilities. These states are very unstable but yet live
a time long enough to be amenable to be treated within stationary
formalisms. We have thus adopted the shell model in the complex energy
plane for this purpose. In addition we performed the shell
model calculation by using the multistep shell model. In this method of
solving the shell model equations one proceeds in several steps. In each
step one constructs building blocks to be used in future steps \cite{lio82}.
We applied this formalism to analyze $^{12,13}$Li as determined by the neutron
degrees of freedom.

By using single-particle energies (i.e., states in $^{10}$Li) as provided by 
experimental data when available or as provided by our calculation, we found 
that the only physically meaningful two-particle states are $^{11}$Li(gs), 
which is a bound state, and $^{11}$Li($2^+_1$), which is a resonance. 
As a result there are only three physically meaningful states
in $^{12}$Li which, besides the antibound ground state, it is predicted that
there is a resonance  $1/2^-$ lying at about 1 MeV and about 800 keV wide
and another resonance which is  $5/2^+$ lying at about 1.1 MeV and 500 keV 
wide. That the ground state is an antibound (or virtual) state was confirmed 
by a number of experiments \cite{aks08,pat10,hal10} and the state $5/2^+$ has
probably been observed in \cite{pat10}. However, in \cite{hal10} two 
additional states, lying at rather low energies, have been observed which do 
not seem to correspond to the calculated levels. It has to be mentioned that 
neither a shell model calculation, performed within an harmonic oscillator 
basis, provides satisfactory results in this case. Yet, we found that this 
shell model calculation works better than one would assume given the 
unstable character of the states involved.
In four-particle system we found that there is no bound or antibound state in $^{13}$Li, which agrees with the experiment that the $^{13}$Li is unbound.

\section*{Acknowledgments}
This work has been supported by the Swedish Research Council (VR) under grant Nos. 623-2009-7340 and 2010-4723.


\medskip
\section*{References}
\begin{thebibliography}{90}
\bibitem{idb02}Betan R Id, Liotta R J, Sandulescu N and Vertse T 2002 {\it Phys. Rev. Lett.} {\bf 89} 042501.
\bibitem{gam28}Gamow G 1928 {\it Z. Phys.} {\bf 51} 204.
\bibitem{mic02}Michel N, Nazarewicz W, Ploszajczak M and Bennaceur K 2002 {\it Phys. Rev. Lett.} {\bf 89} 042502.
\bibitem{cxsm}Betan R Id, Liotta R J, Sandulescu N and Vertse T 2003 {\it Phys. Rev. C} {\bf 67} 014322.
\bibitem{mic09}Michel N, Nazarewicz W, P{\l}oszajczak M and Vertse T 2009 {\it J. Phys. G} {\bf36} 013101.
\bibitem{cxmsm}Xu Z X, Liotta R J, Qi C, Roger T, Roussel-Chomaz P, Savajols H and Wyss R 2011 {\it Nucl. Phys. A} {\bf850} 53 (preprint: arXiv:1012.2303).
\bibitem{blo84}Blomqvist J, Liotta R J, Rydstrom L and Pomar C 1984 {\it Nucl. Phys. A } {\bf 423} 253.
\bibitem{b68} Berggren T 1968 {\it Nucl. Phys. A} {\bf 109} 265.
\bibitem{lio96}Liotta R J, Maglione E, Sandulescu N and Vertse T 1996
{\it Phys. Lett.  B} {\bf 367} 1.
\bibitem{lio82}Liotta R J and Pomar C 1982 {\it Nucl. Phys. A} {\bf 382} 1.
\bibitem{esb97}Esbensen H, BertschG F and Hencken K 1997 {\it Phys. Rev. C} {\bf 56} 3054, and references therein.
\bibitem{idb04}Betan R Id, Liotta R J, Sandulescu N, and Vertse T 2004 {\it Phys. Lett. B} {\bf 584} 48.
\bibitem{ant}Betan R Id, Liotta R J, Sandulescu N, Vertse T and Wyss R 2005 
{\it Phys. Rev. C} {\bf 72} 054322.
\bibitem{aks08}Aksyutina Yu et al. 2008 {\it Phys. Lett. B} {\bf 666} 430.
\bibitem{pat10}Roger T, PhD thesis; Roussel Chomaz P et al., to be published.
\bibitem{hal10}Hall C C et al. 2010 {\it Phys. Rev. C} {\bf 81} 021302(R).
\bibitem{qi08}Qi C and Xu F R 2008 {\it Chin. Phys. C} {\bf 32} (S2) 112.
\end{thebibliography}
\end{document}